\documentclass[journal]{IEEEtran}

\usepackage{booktabs}
\usepackage{xcolor,soul,framed} 
\usepackage{tabularx}
\colorlet{shadecolor}{yellow}
\usepackage[pdftex]{graphicx}
\graphicspath{{../pdf/}{../jpeg/}}
\DeclareGraphicsExtensions{.pdf,.jpeg,.png}
\usepackage{amsfonts} 
\usepackage{amssymb}

\usepackage[cmex10]{amsmath}
\usepackage{array}
\usepackage{mdwmath}
\usepackage{mdwtab}
\usepackage{eqparbox}
\usepackage{url}
\usepackage{comment}
\usepackage{subcaption}
\usepackage{tabularx}
\usepackage{braket}
\usepackage[ruled,vlined,linesnumbered]{algorithm2e}
\usepackage{soul}

\bstctlcite{IEEE:BSTcontrol}

\begin{document}
\bstctlcite{IEEEexample:BSTcontrol}
    \title{Optimizing Low-Energy Carbon IIoT Systems with Quantum Algorithms: Performance Evaluation and Noise Robustness
    }
 
 \author{Kshitij~Dave,
   Nouhaila~Innan,
      Bikash~K.~Behera,
     Shahid~Mumtaz,
     Saif~Al-Kuwari
    and~Ahmed~Farouk

\thanks{K.~Dave is with the Institute of Advanced Research, Gandhinagar, India e-mail: (kshitijdave2@gmail.com).}
\thanks{N.~Innan is Quantum Physics and Magnetism Team, LPMC, Faculty of Sciences Ben M'sick, Hassan II University of Casablanca, Morocco. And with the eBRAIN Lab, Division of Engineering, New York University Abu Dhabi (NYUAD), and the Center for Quantum and Topological Systems (CQTS), NYUAD Research Institute, NYUAD, Abu Dhabi, UAE. e-mail: (nouhaila.innan@nyu.edu).}
\thanks{B.~K. Behera is with the Bikash's Quantum (OPC) Pvt. Ltd., Mohanpur, WB, 741246 India, e-mail: (bikas.riki@gmail.com).}
\thanks{Shahid~Mumtaz is with Nottingham Trent University, Engineering Department, United Kingdom. e-mail: (dr.shahid.mumtaz@ieee.org).}
\thanks{S.~Al-Kuwari is with the Qatar Center for Quantum Computing, College of Science and Engineering, Hamad Bin Khalifa University, Doha, Qatar. e-mail: (smalkuwari@hbku.edu.qa).}
\thanks{A.~Farouk is with the Qatar Center for Quantum Computing, College of Science and Engineering, Hamad Bin Khalifa University, Doha, Qatar and with the Department of Computer Science, Faculty of Computers and Artificial Intelligence, Hurghada University, Hurghada, Egypt. Also, with  e-mail: (ahmedfarouk@ieee.org).}}


\maketitle

\begin{abstract}
Low-energy carbon Internet of Things (IoT) systems are essential for sustainable development, as they reduce carbon emissions while ensuring efficient device performance. Although classical algorithms manage energy efficiency and data processing within these systems, they often face scalability and real-time processing limitations. Quantum algorithms offer a solution to these challenges by delivering faster computations and improved optimization, thereby enhancing both the performance and sustainability of low-energy carbon IoT systems. Therefore, we introduced three quantum algorithms: quantum neural networks utilizing Pennylane (QNN-P), Qiskit (QNN-Q), and hybrid quantum neural networks (QNN-H). These algorithms are applied to two low-energy carbon IoT datasets—room occupancy detection (RODD) and GPS tracker (GPSD). For the RODD dataset, QNN-P achieved the highest accuracy at 0.95, followed by QNN-H at 0.91 and QNN-Q at 0.80. Similarly, for the GPSD dataset, QNN-P attained an accuracy of 0.94, QNN-H 0.87, and QNN-Q 0.74. Furthermore, the robustness of these models is verified against six noise models. The proposed quantum algorithms demonstrate superior computational efficiency and scalability in noisy environments, making them highly suitable for future low-energy carbon IoT systems. These advancements pave the way for more sustainable and efficient IoT infrastructures, significantly minimizing energy consumption while maintaining optimal device performance.
\end{abstract}


%
\IEEEpeerreviewmaketitle


\section{Introduction}



The Internet of Things (IoT) has application domains that enable technologies on energy efficiency techniques for Wireless Sensor Networks (WSNs) and identify gaps for future research on energy preservation measures \cite{Dudhe2021IoT}. The application encompasses analyzing various agriculture domains \cite{farooq2020role}, cloud computing \cite{ihirwe2021cloud}, transportation systems \cite{zantalis2019review}, environment monitoring \cite{islam2020development}, and many more. 
As this technology matured, it expanded beyond consumer applications to more complex and large-scale industrial settings, leading to the emergence of the Industrial Internet of Things (IIoT) \cite{boyes2018industrial,Karmakar2019IoT-IIoT}. IIoT-based applications involve operational efficiency, predictive maintenance, and data-driven decision-making, which ultimately improve productivity and cost efficiency in various industries, including manufacturing, energy, and transportation \cite{raj2021predictive}.
One of the advantages of IIoT is that its enhanced connectivity allows real-time monitoring through WSNs \cite{network1030017}. IIoT also facilitates increased automation, which leads to more decentralized and autonomous manufacturing operations. These capabilities are driving the realization of smart factories in sectors such as utilities, manufacturing, and food production \cite{liao2018industrial}. Although IIoT is transforming industrial landscapes by improving efficiency, safety, and productivity, research indicates that its full potential has not yet been realized, with opportunities for expansion in more industrial sub-sectors \cite{Ahmed2023IIoT}.

Several challenges of IIoT infrastructure include ensuring security in resource-constrained environments, such as protecting against physical attacks, securing communication channels, and maintaining system robustness even under attack. Implementing efficient cryptographic protocols, such as lightweight cryptography and secure identification mechanisms, is difficult due to the high computational overhead \cite{sisinni2018industrial}. 
The growing data volume from heterogeneous IIoT devices poses challenges in data management. Advanced models are needed to efficiently process, transmit, and store large raw datasets, ensuring secure storage, fast retrieval, and high-speed processing to support real-time decision-making in IIoT systems\cite{khan2020industrial}. IIoT systems handle enormous amounts of data while maintaining dependability, security, and energy efficiency. Due to resource constraints, they have limitations, including maximizing energy efficiency and reducing carbon emissions, particularly as the spread of the IIoT raises the world's energy demand. 
Optimization problems such as energy management and carbon reduction are difficult to handle using traditional approaches. 

However, Quantum Computing (QC), particularly Quantum Machine Learning (QML) algorithms, have the potential to overcome these limitations by providing new techniques to reduce carbon footprints and increase energy efficiency in IIoT systems \cite{rahman2021quantum,smith2020energy}.
\begin{figure*}
\centering
\includegraphics[width=1\linewidth]{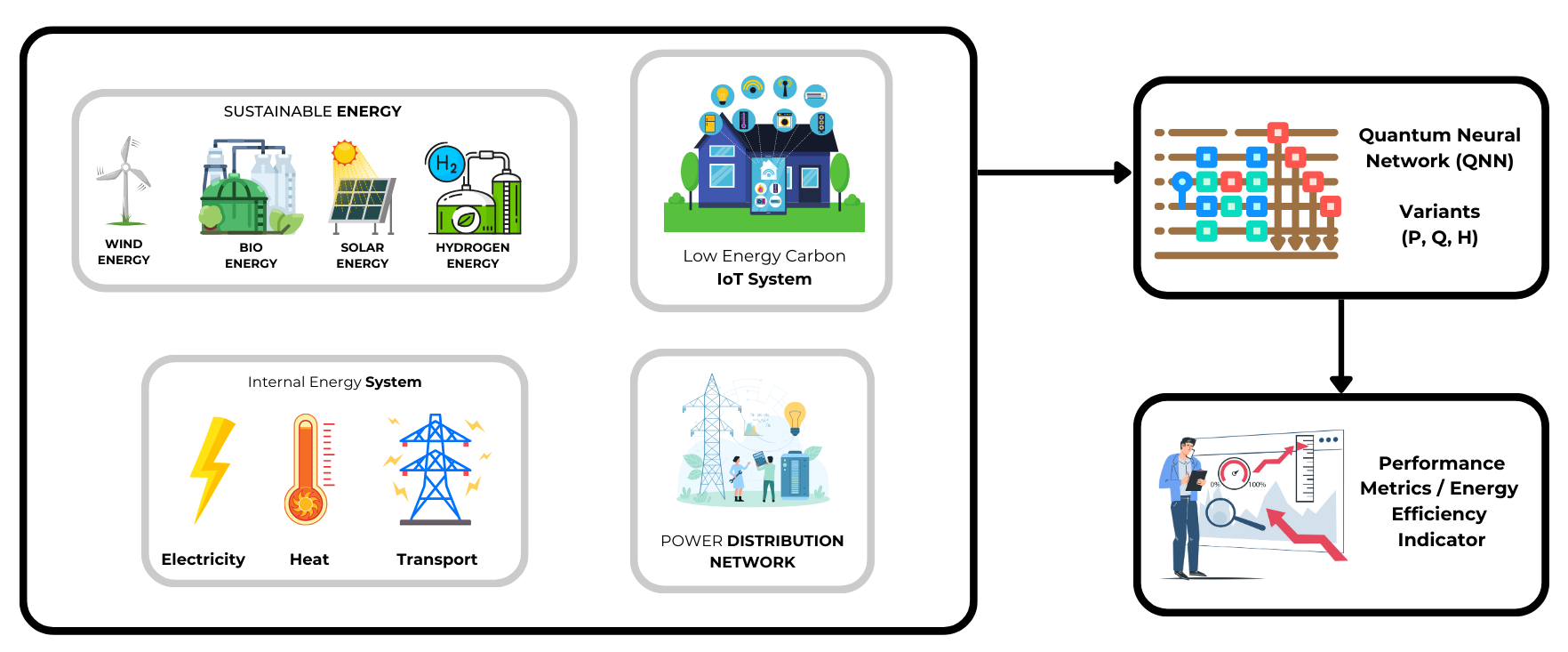}
\caption{Integration of QNN Algorithms with Low-Energy Carbon IIoT Systems for Optimized Energy Management.}
\label{fig:Schematic}
\end{figure*}
A technique inspired by QC is introduced to optimize temporal space in real-time IIoT applications, focusing on accuracy, energy efficiency, and data similarity analysis \cite{bhatia2023novel}. A novel GAN-QEHO algorithm, which combines adversarial generative networks (GAN) with the optimization of the quantum elephant herd (QEHO), is presented to improve mobile edge computing (MEC) in big data environments enabled by IIoT \cite{kaur2022generative}. Quantum algorithms, such as annealing, genetic algorithms, and particle swarm optimization, solve resource allocation, network routing, and energy efficiency challenges in IIoT networks.
\cite{raparthi2022quantum}. The problem of embedding service function chains (SFC) in resource-demanding IIoT networks is tackled by reformulating it into a quadratic unconstrained binary optimization formulation. A hybrid warm-start quantum annealing technique is proposed to improve resource utilization, computational speed, and scalability in deploying SFC \cite{emu2022resource}. QML algorithms are shown to significantly improve the prediction of power generation in extreme environments of the IIoT, achieving higher accuracy than classical methods, which could improve energy efficiency and decision-making \cite{satpathy2023analysis}.


While numerous quantum algorithms are applied in IIoT systems, there remains potential to enhance their efficiency further. Additionally, the behavior of these algorithms under the influence of noise channels has not yet been fully explored. To address these gaps, we propose novel quantum algorithms to improve efficiency and demonstrate robustness against various noise models (see Fig.\ref{fig:Schematic}). Our study implements quantum neural networks using the Qiskit and Pennylane frameworks, specifically focusing on QNN-Q (Qiskit), QNN-P (Pennylane), and QNN-H (Hybrid QNN). We examine the impact of different noise models on the accuracy of the algorithms as noise parameters increase. The results are then compared with previous studies to assess performance improvements and validate the effectiveness of our proposed models.

The main contributions of this work are summarized as follows:
\begin{itemize}
    \item[1)] Three quantum algorithms—QNN-P, QNN-Q, and QNN-H—are introduced for optimizing low-energy carbon IIoT systems.
    \item[2)] The superiority of these algorithms is demonstrated through extensive simulation experiments on two low-energy carbon datasets, with their performances compared using key evaluation metrics.
    \item[3)] The robustness of the best-performing algorithm is validated against six different noise models, showing enhanced resilience in noisy environments.
\end{itemize}
The paper is structured as follows: Section \ref{SecIII} provides an overview of the framework and problem formulation. Section \ref{SecIV} discusses experimental settings and the results for the RODD and GPSD datasets, along with an analysis of noise effects. Finally, Section \ref{SecV} concludes the paper with a summary of the results and key insights.
\begin{figure}[]
    \centering
    \begin{subfigure}{0.5\textwidth}
        \centering
        \includegraphics[width= \linewidth]{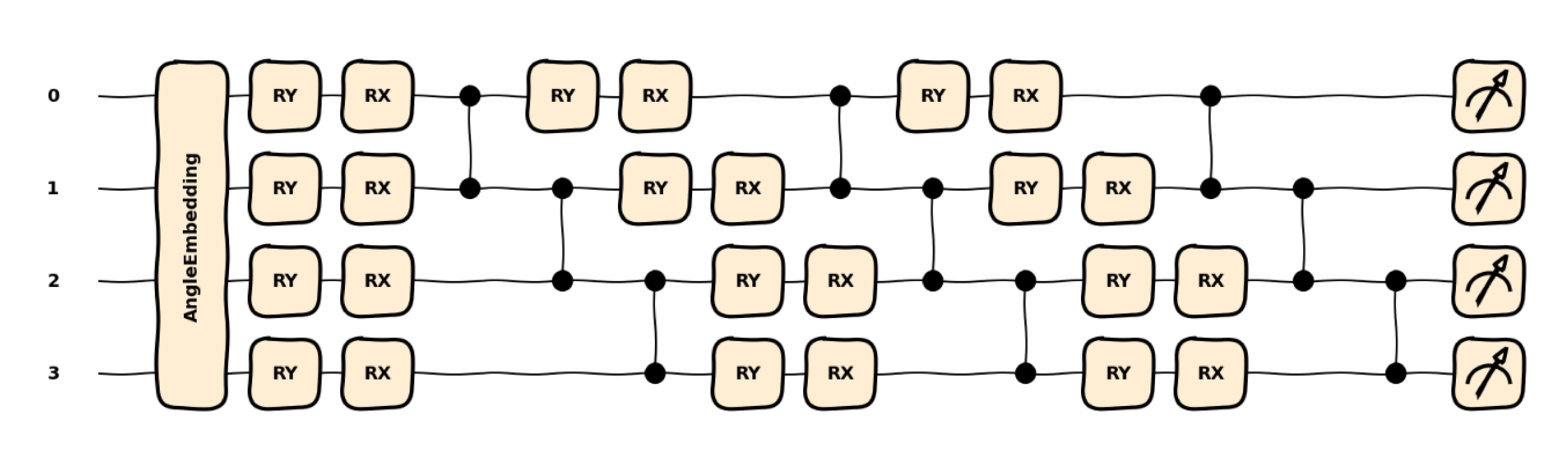}
        \caption{}
        \label{fig:QNN-P}
    \end{subfigure}\hfill
    \begin{subfigure}{0.5\textwidth}
        \centering
        \includegraphics[width=\linewidth]{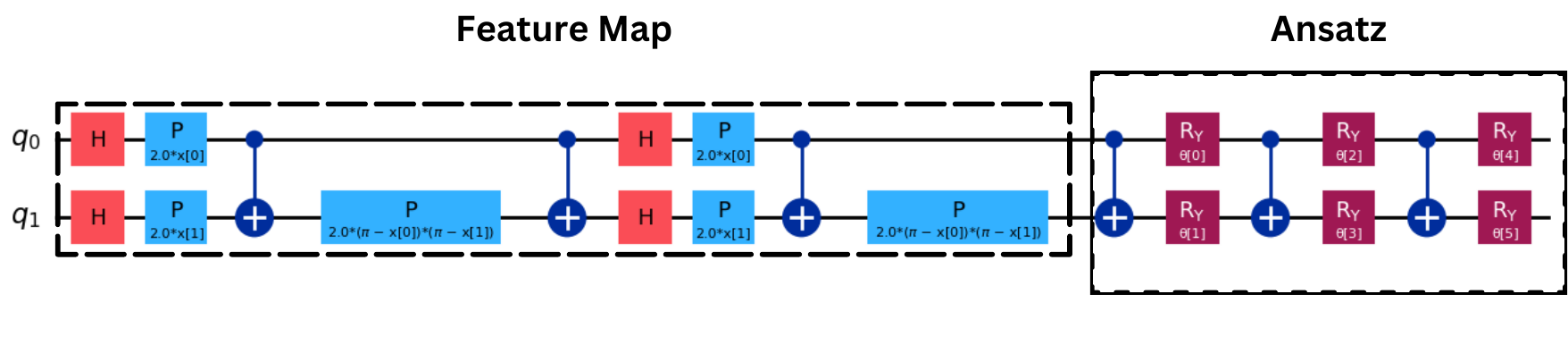}
        \caption{}
        \label{fig:QNN-Q}
    \end{subfigure}\hfill 
    \begin{subfigure}{0.5\textwidth}
        \centering
        \includegraphics[width=\linewidth]{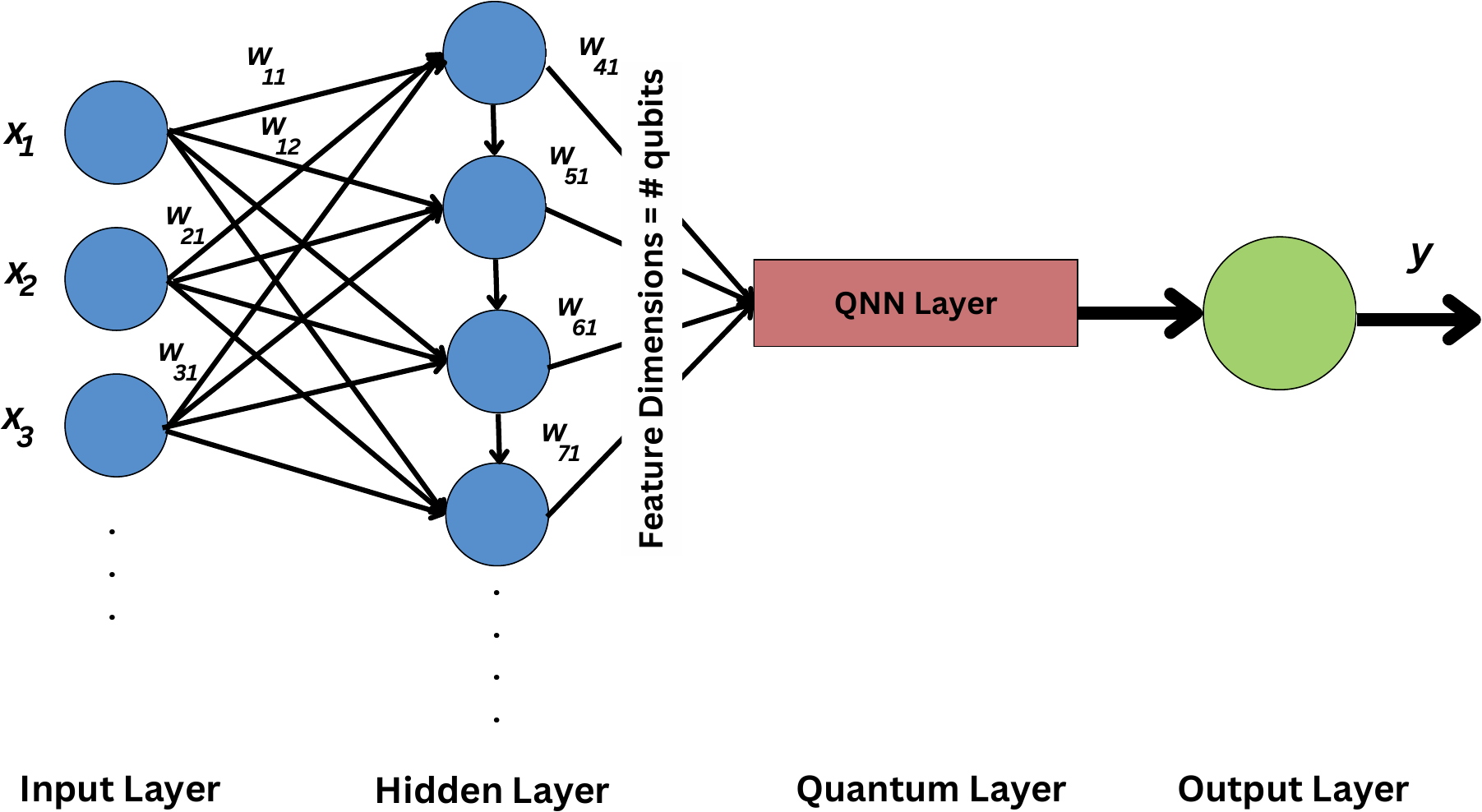}
        \caption{}
        \label{fig:qnn-H}
    \end{subfigure}\hfill
    \caption{Representation of quantum models: (a) The QNN-P architecture utilizes angle embedding for feature encoding, followed by $R_Y$ and $R_X$  rotation gates and entanglement operations, (b) The QNN-Q architecture employs a ZFeatureMap for data embedding and a RealAmplitudes ansatz to construct the quantum circuit, and (c) The QNN-H architecture comprises an input layer for classical feature encoding, a hidden layer, and a quantum layer that integrates a feature map and ansatz, culminating in an output layer for prediction.}
    \label{Fig5a-}
\end{figure}
\section{Methodology \label{SecIII}}

\subsection{Overall Framework and Problem Formulation}

Here, we present the theoretical framework and problem formulation that underpin the experimental analysis of two datasets: RODD and GPSD. These datasets introduce data size and complexity challenges, offering realistic scenarios reflective of real-world IoT applications. Our framework is designed to identify the optimal set of parameters \( \theta_i \), minimizing loss through hyperparameter tuning while ensuring model robustness in noisy environments.
Mathematically, it can be written as:-

\begin{equation}
\min_{\theta_i} \quad \mathcal{L}(\hat{y_i}, y_i) = \min_{\theta_i} \quad \mathcal{L}(f(X; \theta_i), y_i),
\end{equation}

where, $\hat{y_i} = f(X; \theta_i)$ represents the predicted output, and $y_{i}$ is the true label.
$\mathcal{L}$ is the loss function, and $\theta_i$ are optimized parameters.
Typical, loss function uses binary cross entropy or mean square error in classification. The parameterized quantum circuit in the form of unitary operations $U(\theta)$ is applied to an initialized statevector $\ket{\psi_0}$ to produce final quantum circuit $\ket{\psi(\theta)}$ shown as:-

\begin{equation}
|\psi(\theta)\rangle = \prod_{i=1}^{n} U(\theta_{i})|\psi_0\rangle.
\end{equation}
 To assess the model's performance, we focus on maximizing key metrics which are formally defined as follows:

\begin{eqnarray}
&&\max_{\theta} \quad \text{Precision}(\hat{y}, y), \quad \text{Recall}(\hat{y}, y),\nonumber\\
&&\quad \text{F1\_Score}(\hat{y}, y), \quad \text{Accuracy}(\hat{y}, y).
\end{eqnarray}
The experiment systematically explores the capabilities of quantum and hybrid models by varying circuit complexity by modifying the feature maps, the depth of the ansatz, and the configuration of entangling gates. The quantum circuit design is hypothesized to affect model efficiency and accuracy, with feature maps encoding classical data and the ansatz representing the solution space through parameterized operations. The goal is to optimize performance by fine-tuning these elements.

\begin{equation}
\text{Feature Map Configuration} = \sum_{i=1}^{n} \text{FM}(x_i; \phi_i),
\end{equation}

\begin{equation}
\text{Ansatz Configuration} = \prod_{k=1}^{r} U_k(\theta_k).
\end{equation}



\subsection{Ansatz}
\subsubsection{RealAmplitudes}
The RealAmplitudes ansatz consists of alternating layers of single-qubit rotation and two-qubit (entangling) gates.
Each qubit undergoes a rotation around the Y-axis:
\begin{equation}
    R_{y}(\theta) = 
    \begin{bmatrix}
    \cos{\theta/2} & -\sin{\theta/2} \\
    \sin{\theta/2} & \cos{\theta/2}
    \end{bmatrix}.
\end{equation}
The entangling gates are typically controlled-NOT (CNOT) gates applied between pairs of qubits; they are defined as,

\begin{equation}
CNOT_{i,j} = \ket{0}\bra{0} \otimes I + \ket{1}\bra{1} \otimes X
\end{equation}

The overall ansatz $n$ qubits and $r$ repetitions, the ansatz can be defined as follows:
\begin{equation}
    U(\theta) = \prod_{k=1}^r \left( \bigotimes_{i=1}^n R_{y}(\theta_{i,k}) \right) \left( \prod_{(i,j) \in E} \text{CNOT}_{i,j} \right),
\end{equation}
Where $E$ represents the set of qubit pairs for two-qubit gates.

\subsubsection{EfficientSU2 Ansatz}
`EfficientSU2' circuit consists of layers of single-qubit operations that cover the SU(2) space with minimal parameters.

\begin{enumerate}
  \item Single-Qubit Rotations:
    $R_y(\theta)R_z(\phi)$
  \item Entangling Gates:-
    Commonly used CNOT gates to entangle between qubits:
    $CNOT_{i,i+1}$
\end{enumerate}
The ansatz can be expressed as:
\begin{equation}
    U(\theta,\phi) = \prod_{k=1}^{r} \left( \bigotimes_{i=1}^{n} R_{y}(\theta_{i,k}, \phi_{i,k}) \right) \left( \prod_{(i,j) \in E} \text{some operation} \right).
\end{equation}
The tensor product is used to apply each rotational gate independently to its corresponding circuit depth.
\subsection{Feature Map}

\subsubsection{ZFeatureMap}

It is defined as an alternative layer of Hadamard gates (H) and phase gate $P(2x_{i})$ gates. The phase gate encodes classical input data into a quantum state by rotating each qubit around the Z-axis with angles proportional to $2x_{i}$.

Quantum gates used:
\begin{enumerate}
    \item Hadamard gate (H): $H = \frac{1}{\sqrt{2}} \begin{bmatrix}
        1 & 1\\
        1 & -1
    \end{bmatrix}.$

    \item Phase gate (P): $P(\theta) = \begin{bmatrix}
        1 & 0\\
        0 & e^{i\theta}
    \end{bmatrix}.$ 
\end{enumerate}
The overall equation can be written for $n$ qubits, $r$ repetitions, and the input vector $x_{i}$ as follows:
\begin{equation}
    \phi(x) = \prod_{k=1}^{r} \left( \bigotimes_{i=0}^{n-1} H \right) \left(\bigotimes_{i=0}^{n-1} P(2x_{i}) \right).
\end{equation}
There are no entangling gates in this feature map, which generally gives an advantage when the complexity of input data is not complex.

\subsubsection{ZZFeatureMap}
The overall feature map representation of $n$ qubits, $r$ repetitions, and $x_{i},x_{j}$ for input data components.

\begin{align}
    \Phi(x) &= \prod_{k=1}^r \left( \left( \bigotimes_{i=0}^{n-1} H \right) \left( \bigotimes_{i=0}^{n-1} P(2 x_i) \right) \right.\nonumber\\
    &\quad \left. \left( \prod_{(i,j) \in E} \left( CNOT_{ij} \cdot P(2 x_i x_j) \cdot CNOT_{ij} \right) \right) \right),
\end{align}

where $E$ represents the set of pairs of qubits to be entangled and $P(2x_{i}x_{j})$ is the phase gate applied before the first and second CNOT gates.

\subsubsection\hl{{Pennylane Embedding:}}

Data embedding into the quantum state is performed through angle embedding. For a normalized data vector $x \in \mathbb{R}^n$, the quantum state after embedding is defined as:
\begin{equation}
|\Psi_{\text{data}}(x)\rangle = \bigotimes_{i=1}^m R_Y(x_i) |0\rangle.
\end{equation}

The network incorporates three sequential layers, each composed of parametric rotational gates $R_Y(\theta)$ and $R_X(\phi)$, applied to each of the four qubits. The unitary operation for each qubit $j$ in layer $k$ is:
\begin{equation}
U_j^{(k)}(\theta_j^{(k)}, \phi_j^{(k)}) = R_X(\phi_j^{(k)}) R_Y(\theta_j^{(k)}).
\end{equation}
where, $\phi_j^{(k)}$ and $\theta_j^{(k)}$ are trainable parameters.
Following the rotations, CNOT gates are systematically applied to entangle the qubits, enhancing their ability to capture correlations in the data.

\subsection{QNN-P}
\begin{algorithm}[!h]
\DontPrintSemicolon
\caption{QNN-P}
\label{alg:qnn-p}
\KwIn{Training dataset $X_{train}$ and labels $y_{train}$, test dataset $X_{test}$ and labels $y_{test}$, maximum iterations, learning rate.}
\KwOut{Trained QNN-P}

\textbf{Initialize:} Quantum device with 4 qubits\;
Define QNN structure with 3 layers of $R_Y$, $R_X$ rotations and CNOT gates\;
Define quantum node \textit{circuit} using Angle Embedding and QNN\;
Normalize and preprocess $X_{train}$ to $X_{train\_demo\_flattened}$\;
Initialize empty lists for loss and accuracy history\;

\For{$iteration\gets1$ \KwTo $max\_iterations$}{
    Initialize QNN parameters $params$ with random values\;
    Define cost function as MSE between QNN predictions and true labels\;
    Update $params$ using gradient descent with Adam optimizer\;
    Append current loss to loss history\;
    \For{each data point in $X_{train\_demo\_flattened}$}{
        Predict label using a \textit{circuit} with current $params$\;
    }
    Compute accuracy and append to accuracy history\;
}

\textbf{Testing Phase:}\;
Normalize and preprocess $X_{test}$ to $X_{test\_flattened}$\;
\For{each data point in $X_{test\_flattened}$}{
    Predict label using a \textit{circuit} with optimized $params$\;
}
Calculate final accuracy on $X_{test}$\;
Plot loss and accuracy history\;

\Return{Optimized parameters $params$, training loss and accuracy history, performance metrics, predictions}\;

\end{algorithm}

Each qubit in the four-qubit system is initialized in the state $|0\rangle$ (Fig. \ref{fig:QNN-P}). The full quantum system state is:
\begin{equation}
|\Psi_{\text{init}}\rangle = |0\rangle^{\otimes 4}.
\end{equation}

Data embedding into the quantum state is achieved through angle embedding. Given a normalized data vector $x \in \mathbb{R}^n$, the quantum state after embedding is defined as:
\begin{equation}
|\Psi_{\text{data}}(x)\rangle = \bigotimes_{i=1}^m R_Y(x_i) |0\rangle.
\end{equation}

The network incorporates three sequential layers, each composed of parametric rotational gates $R_Y(\theta)$ and $R_X(\phi)$, applied to each of the four qubits. The unitary operation for each qubit $j$ in layer $k$ is:
\begin{equation}
U_j^{(k)}(\theta_j^{(k)}, \phi_j^{(k)}) = R_X(\phi_j^{(k)}) R_Y(\theta_j^{(k)}).
\end{equation}
Following the rotations, CNOT gates are systematically applied to entangle the qubits, enhancing their ability to capture correlations in the data. The outcome of the quantum circuit is measured using the Pauli-Z observable on the first qubit. The expected value of this measurement, which directly correlates with the prediction output, is given by:
\begin{equation}
\langle Z_0 \rangle = \langle \Psi_{\text{out}} | Z_0 | \Psi_{\text{out}} \rangle,
\label{Eq17}
\end{equation}
where $|\Psi_{\text{out}}\rangle$ represents the quantum state post-application of the circuit operations. The mean squared error (MSE) between the predicted outputs \(\hat{y}\) and the true labels \(y\) is used as the loss function, quantitatively described by:
\begin{equation}
\mathcal{L}(\theta, \phi) = \frac{1}{N} \sum_{i=1}^{N} (\langle Z_0^{(i)} \rangle - y_i)^2,
\label{Eq18}
\end{equation}
where $\langle Z_0^{(i)} \rangle$ is the expected measurement result for the $i$-th data point, and $y_i$ is the corresponding label. The parameters $\theta$ and $\phi$ in the quantum circuit are optimized by the Adam optimizer, an advanced variant of stochastic gradient descent. Adam adapts the learning rates for each parameter by leveraging exponentially weighted moving averages of the gradients, making it particularly effective in dynamically adjusting learning rates throughout the optimization process. 
The gradient of the loss function for each parameter $\theta_j $ and $\phi_j $ is computed using the parameter-shift rule:
\begin{equation}
\frac{\partial \mathcal{L}}{\partial \theta_j} = \frac{\mathcal{L}(\theta_j + \pi/2) - \mathcal{L}(\theta_j - \pi/2)}{2},
\label{Eq19}
\end{equation}
and similarly for $\phi_j$. The parameter-shift rule is utilized here due to its ability to provide exact gradients for quantum gates with parameterized rotations, which is crucial for effective learning in QNN (see Algorithm \ref{alg:qnn-p}). The Adam optimizer updates the parameters according to the following equation:
\begin{equation}
\theta_j^{(t+1)} = \theta_j^{(t)} - \eta \frac{\hat{m}_t}{\sqrt{\hat{v}_t} + \epsilon},
\label{Eq19}
\end{equation}
where $\eta$ is the learning rate, $\hat{m}_t$ and $\hat{v}_t$ are bias-corrected estimates of the first and second moments, and $\epsilon$ prevents division by zero. The moment estimates are calculated as:
\begin{eqnarray}
m_t &=& \beta_1 m_{t-1} + (1 - \beta_1) g_t, \quad \hat{m}_t = \frac{m_t}{1 - \beta_1^t},\nonumber\\
v_t &=& \beta_2 v_{t-1} + (1 - \beta_2) g_t^2, \quad \hat{v}_t = \frac{v_t}{1 - \beta_2^t},
\label{Eq20}
\end{eqnarray}
where $g_t$ represents the gradient of the loss function with respect to the parameters at step \( t \), and $\beta_1$ and $\beta_2$ are the exponential decay rates for the moment estimates. 
The training process uses the classical optimizer to update model parameters iteratively over multiple epochs. Each epoch includes a forward pass to calculate loss, followed by a backward pass to update parameters based on gradients. The optimized parameters are then applied to unseen data for model evaluation, ensuring effective generalization.

\subsection{QNN-Q}
\begin{algorithm}[!h]
\DontPrintSemicolon
\caption{QNN-Q}
\label{alg:qnn-q}
\KwIn{Training dataset $X_{train}$ and labels $y_{train}$, test dataset $X_{test}$ and labels $y_{test}$, maximum iterations, learning rate.}
\KwOut{Trained QNN-Q}

\textbf{Initialize:} Quantum device with 2 qubits\;
Define QNN structure with feature map and ansatz with input dimension and suitable repetition value\;
Define EstimatorQNN module to set QNN\;
Randomly select inputs for training and testing for the target feature such that each subset contains an equal number of 0s and 1s.\;
Normalize and preprocess defined subset $X$ and $y$\;
Initialize loss function $BCEWithLogitsLoss()$ and optimizer $Adam$.
Initialize empty lists for loss and accuracy history\;
Initialize model for training\;
\For{$iteration\gets1$ \KwTo epochs}{
    define empty array predicted and correct labels\;
    \For{each data point in $X$, $y$}{    
    Compute the model output. A forward pass of the input through the model.\;
    Calculate loss function.\;
    append correct \& calculated predicted labels.\;
    clear the gradients.\;
    Backpropagating the error through the model.\;
    Update the optimizer to reduce the loss.\;    
}
Compute accuracy and append loss and accuracy.\;
}

\textbf{Testing Phase:}\;
Put the model in evaluate mode.\;
Use inference mode.\;
Define array for accuracy and loss to store.\;
\For{each data point in $X$,$y$ in test dataset}{
     Compute the model output. \;
     Calculate loss function.\;
     Append the $y$ and logits output in the array.\;
}
Calculate final accuracy on $X_{test}$\;
Plot loss and accuracy history\;

\Return{Optimized parameters $params$, training loss and accuracy history, performance metrics, predictions}\;

\end{algorithm}
The QNN structure is built using two 2-qubit quantum circuits, incorporating a feature map and an ansatz (Fig. \ref{fig:QNN-Q}). The `EstimatorQNN' module is used to develop the QNN-Q algorithm. During preprocessing, balanced data points are selected for training and testing. The loss function employed is `BCEWithLogitsLoss,' and the Adam optimizer updates the model parameters. During training, it calculates outputs and loss, updates gradients and records accuracy at each epoch as shown in Algorithm\ref{alg:qnn-q}. In the testing phase, the model operates in evaluation mode to compute final accuracy and loss see Eqs. \ref{Eq17}-\ref{Eq20}.

\subsection{QNN-H}
\begin{algorithm}[!h]
\DontPrintSemicolon
\caption{QNN-H}
\label{alg:qnn-h}
\KwIn{Training dataset $X_{train}$ and labels $y_{train}$, test dataset $X_{test}$ and labels $y_{test}$, maximum iterations, learning rate}
\KwOut{Trained QNN-Q}

\textbf{Initialize:} Quantum device with 2 qubits\;
Define QNN structure with feature map and ansatz with input dimensions and suitable repetitions value\;
Define EstimatorQNN module to set QNN\;
Define hybrid model using TorchConnector with QNN and neural network integration\;
Randomly select inputs for training and testing for target feature such that each subset contains an equal number of 0s and 1s\;
Normalize and preprocess defined subset $X$ and $y$ \;
Initialize loss function $BCEWithLogitsLoss()$ and $Adam$ optimizer\;
Initialize empty lists for loss and accuracy history\;
Initialize model for training\;
\For{$iteration\gets1$ \KwTo epochs}{
    define empty array predicted and correct labels\;  
    \For{each data point in $X$, $y$}{  
    Compute the model output. A forward pass of the input through the model\;
    Calculate loss function\;
    append correct \& calculated predicted labels\;
    clear the gradients.\;
    Backpropagating the error through the model\;
    Update the optimizer to reduce the loss\;
}
Compute accuracy and append loss and accuracy\;
}

\textbf{Testing Phase:}\;
Put the model in evaluate mode.\;
Use inference mode.\;
Define array for accuracy and loss to store.\;
\For{each data point in $X$,$y$ in test dataset}{
     Compute the model output. \;
     Calculate loss function.\;
     Append the $y$ and logits output in the array.\;
}
Calculate final accuracy on $X_{test}$\;
Plot loss and accuracy history\;

\Return{Optimized parameters $params$, training loss and accuracy history, performance metrics, predictions}\;
\end{algorithm}
The QNN-H algorithm uses Qiskit syntax and integrates a CNN with a QNN via the TorchConnector, forming a hybrid model that combines quantum and classical computational strengths (Fig. \ref{fig:qnn-H}). The initialization phase mirrors QNN-Q's but includes defining the hybrid model. The training process involves forward passes, loss calculation, and backpropagation, similar to QNN-Q but with enhanced computational power due to the hybrid approach, as described in Algorithm \ref{alg:qnn-h}. The testing phase evaluates the model on the test dataset and aims for superior performance in classification tasks by leveraging both quantum and classical ML. The QNN-H algorithm is evaluated using Eqs. \ref{Eq17}-\ref{Eq20}.

\subsection{Noise Model}
Here, the proposed algorithms are rigorously benchmarked in the presence of six different realistic noise models\cite{satpathy2023analysis}. This includes bitflip, phaseflip, and bitphaseflip noise models, which represent fundamental unitary errors arising from gate imperfections or environmental interactions. Then depolarizing noise, a general error model where qubits experience any of the three Pauli errors with equal probability, reflects random and unbiased environmental perturbations. Phase damping and amplitude damping, which are non-unitary noise models capturing essential decoherence effects such as dephasing and energy loss, are crucial for simulating realistic quantum system behavior. This comprehensive evaluation, utilizing these six noise models, demonstrates the robustness and performance of our algorithm against both discrete and continuous noise errors, adhering to standard QC research practices.

The explanation and Kraus operators of the noise models are as follows:

\subsubsection{Bit flip Model}
\begin{equation} 
    E_0 = \sqrt{1 - \eta_B}I , \ E_1 = \sqrt{\eta_B} X
\end{equation}
Here, $\eta_B$ is the probability of a bit-flip error occurring. This error flips the state $\ket{0}$ to $\ket{1}$ and vice-versa. $E_0$ shows qubits remain in the original state, with probability $\sqrt{1 -\eta_B}$, and $E_1$ shows a bit-flip, with probability $\sqrt{\eta_B}$. The overall effect on the state is:
\begin{equation}
    \zeta = (1 - \eta_B)\rho + \eta_B X \rho X
\end{equation}
\subsubsection{Phase-Flip Noise:}
\begin{equation}
    E_0 = \sqrt{1 - \eta_p}I , \ E_1 = \sqrt{\eta_p} Z
\end{equation}
Here, $\eta_P$ probability of occurring a phase-flip error. $E_0$ keeps the qubit in its original phase with probability $1- \eta_p$, and $E_1:$ flips the phase, changing $\ket{+}$ to $\ket{-}$, with probability $\eta_p$. The overall effect on the state is:
\begin{equation}
    \zeta = (1 - \eta_p)\rho + \eta_p Z\rho Z
\end{equation}
\subsubsection{Bit-phase Flip Noise}
\begin{equation}
    E_0 = \sqrt{1 - \eta_{BP}} I , \
    E_1 = \sqrt{\eta_{BP}} Y
\end{equation}
Here, $\eta_{BP}$ is the probability of occurring combined phase and bit flip error. $E_0$ represents no change with probability $1- \eta_{BP}$, and $E_1$ applies both phase and bit flip with probability $\eta_{BP}$. The overall effect on the state is:
\begin{equation}
    \zeta = (1 - \eta_BP)\rho + \eta_BP Y_\rho Y
\end{equation}
\subsubsection{Depolarizing Noise}
\begin{eqnarray}
    E_0 &=& \sqrt{1 - \eta_D} I, E_1 = \sqrt{\eta_D / 3} X,\nonumber\\
    E_2 &=& \sqrt{\eta_D / 3} Y, E_3 = \sqrt{\eta_D / 3} Z
\end{eqnarray}
Here, $\eta_D$ is the probability of occurring depolarizing error. $E_0$ leaves qubit unchanged with probability $1- \eta_D$, and $E_1$, $E_2$, $E_3$ apply the Pauli-X, Pauli-Y, and Pauli-Z operations, respectively with equal probability $\eta_D/3$. The overall effect on $\rho$ is:
\begin{equation}
    \zeta = (1 - \eta_D)\rho + \eta_D/3 (X\rho X + Y\rho Y + Z\rho Z)
\end{equation}
\subsubsection{Phase Damping Noise}
\begin{equation}
    E_0 = \ket{0}\bra{0}+\sqrt{1 - \lambda_P}\ket{1}\bra{1}, \ 
    E_1 = \sqrt{\lambda_P}\ket{1}\bra{1}
\end{equation}
Here, $\lambda_P:$ The probability of occurring depolarizing error. $E_0$ leaves qubit in the original state with probability $1- \lambda_P$, and $E_1$ applies noise between the qubit states with probability $\lambda_P$. The overall effect on $\rho$ is:
\begin{equation}
    \zeta = E_0 \rho E_0^\dagger + E_1 \rho E_1^\dagger
\end{equation}
\subsubsection{Amplitude Damping Noise}
\begin{equation}
    E_0 = \sqrt{1 - \beta_A}\ket{1}\bra{1}, \ E_1 = \sqrt{\beta_A}\ket{0}\bra{1}
\end{equation}
where, $\lambda_A$ is the probability of occurring amplitude damping. $E_0$ leaves qubit in the original state with probability $1- \beta_A$, and $E_1$ applies noise between the qubit states with probability $\beta_A$. The overall effect on $\rho$ is:
\begin{equation}
    \zeta = E_0 \rho E_0^\dagger + E_1 \rho E_1^\dagger
\end{equation}
\section{Experimental Settings and Results \label{SecIV}}
\begin{figure*}[htbp]
    \centering
    \begin{subfigure}{.33\textwidth}
        \includegraphics[width=1\linewidth]{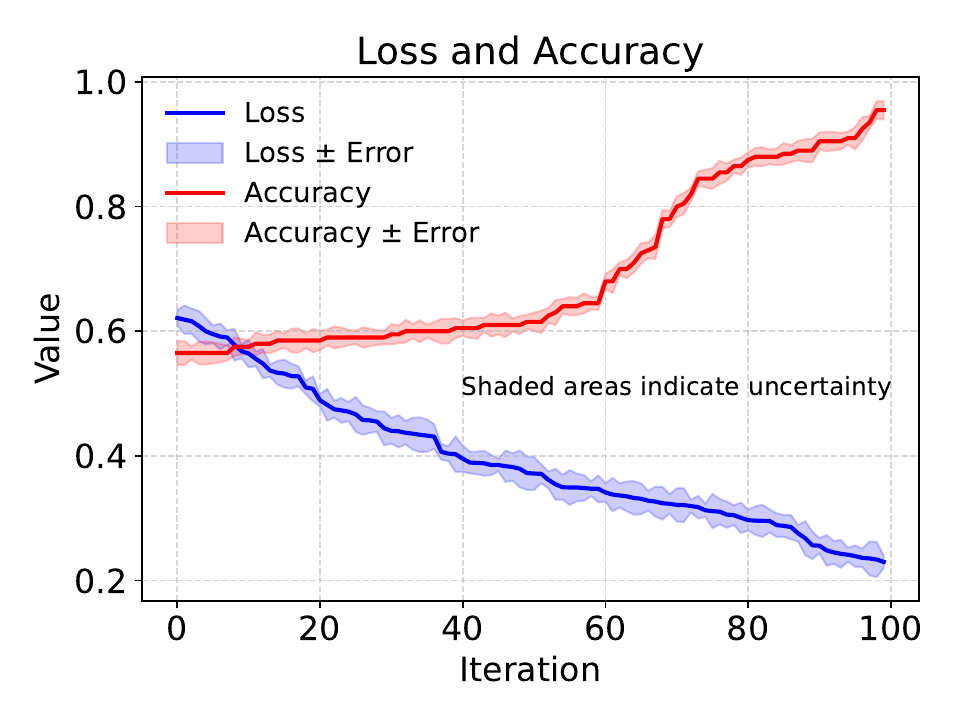}
        \caption{}
        \label{fig:qnn1-P}
    \end{subfigure}\hfill
    \begin{subfigure}{.33\textwidth}
        \includegraphics[width=1\linewidth]{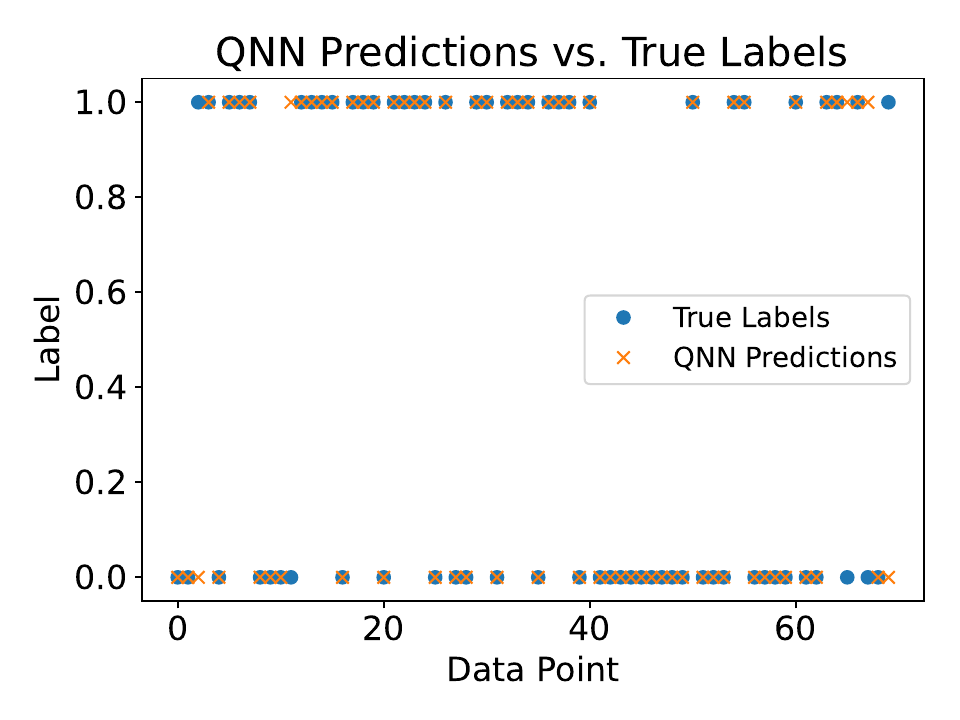}
        \caption{}
        \label{fig:qnn1}
    \end{subfigure}
    \begin{subfigure}{.33\textwidth}
        \includegraphics[width=1\linewidth]{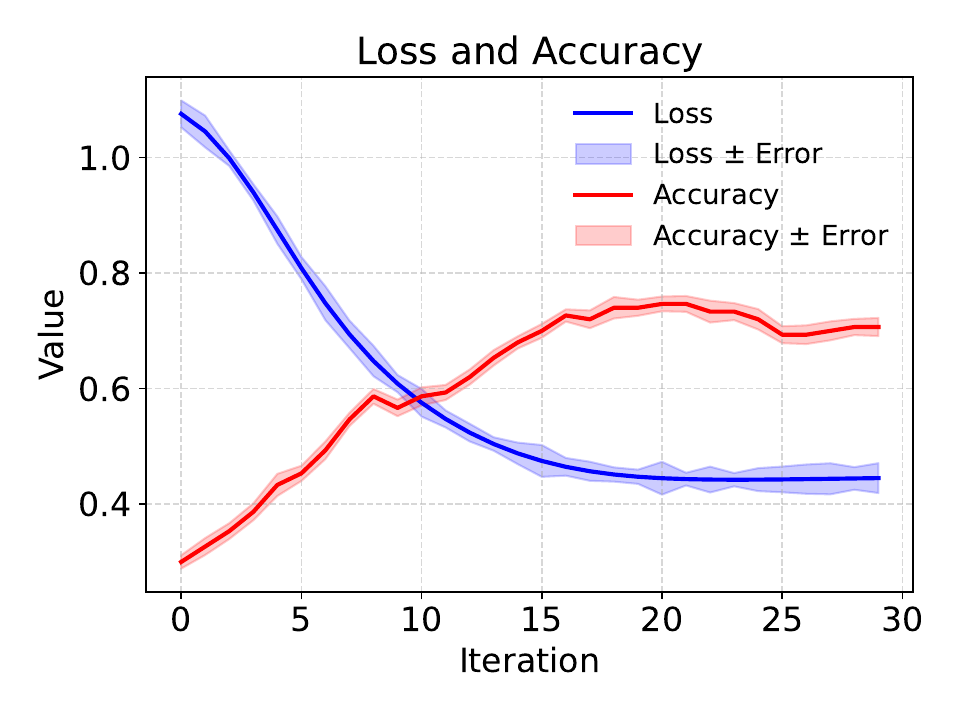}
        \caption{}
        \label{fig:occupancyqnnqiskit}
    \end{subfigure}\hfill
    \begin{subfigure}{.33\textwidth}
        \includegraphics[width=1\linewidth]{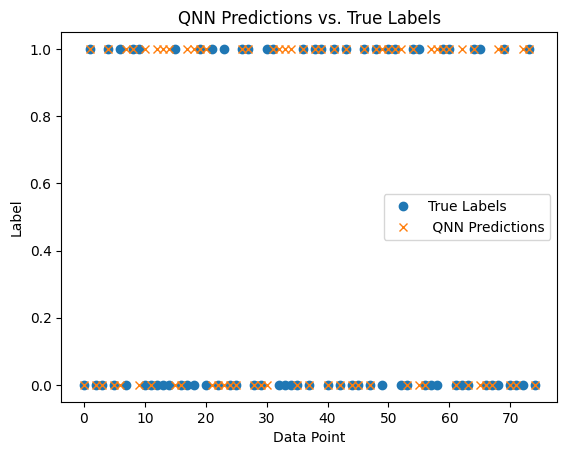}
        \caption{}
        \label{fig:occupancyqnnqiskitdatapoints}
    \end{subfigure}
    \begin{subfigure}{.33\textwidth}
        \includegraphics[width=1\linewidth]{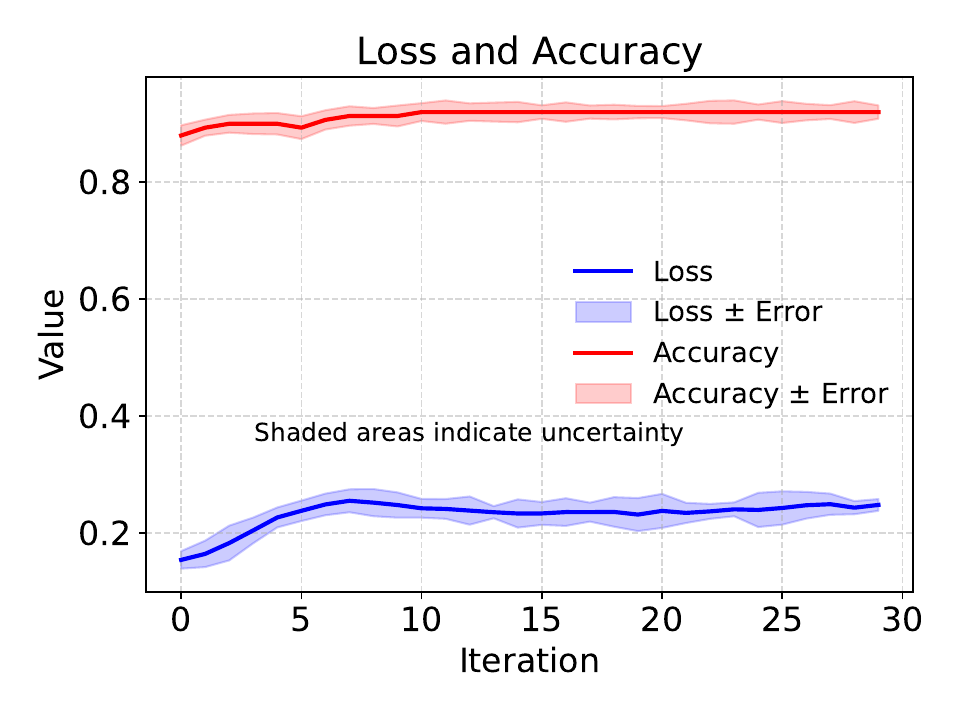}
        \caption{}
        \label{fig:occupancyhqnnqiskit}
    \end{subfigure}\hfill
    \begin{subfigure}{.33\textwidth}
        \includegraphics[width=1\linewidth]{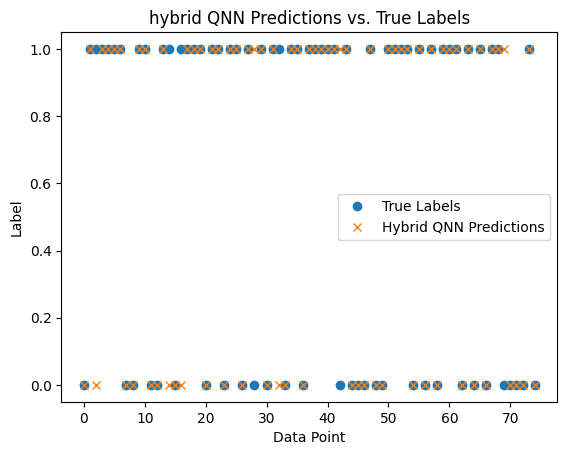}
        \caption{}
        \label{fig:occupancyhqnnqiskitdatapoints}
    \end{subfigure}   
    \caption{Comparative results of different QNN approaches for RODD. Loss and Accuracy curves using (a) QNN-P, (c) QNN-Q, and (e) QNN-H. Predictions using (b) QNN-P, (d) QNN-Q, and (f) QNN-H.}
    \label{Fig4-}
\end{figure*}

\begin{figure*}[htbp]
    \centering
    \begin{subfigure}{.33\textwidth}
        \includegraphics[width=\linewidth]{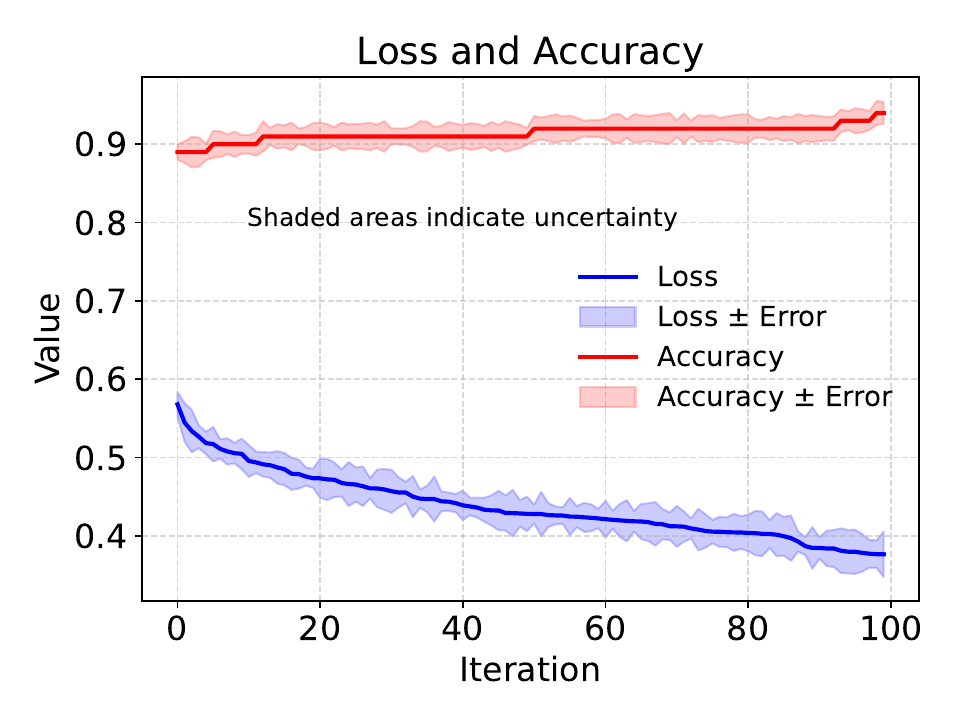}
        \caption{}
        \label{fig:qnn2-P}
    \end{subfigure}\hfill
    \begin{subfigure}{.33\textwidth}
        \includegraphics[width=\linewidth]{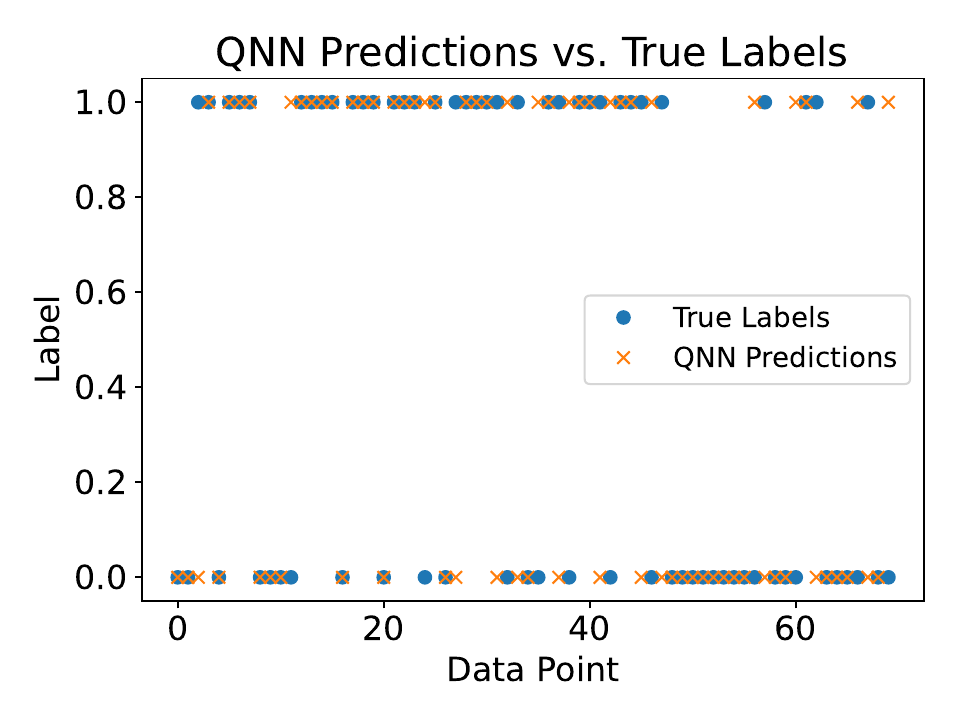}
        \caption{}
        \label{fig:qnn2}
    \end{subfigure}
    \begin{subfigure}{.33\textwidth}
        \includegraphics[width=1\linewidth]{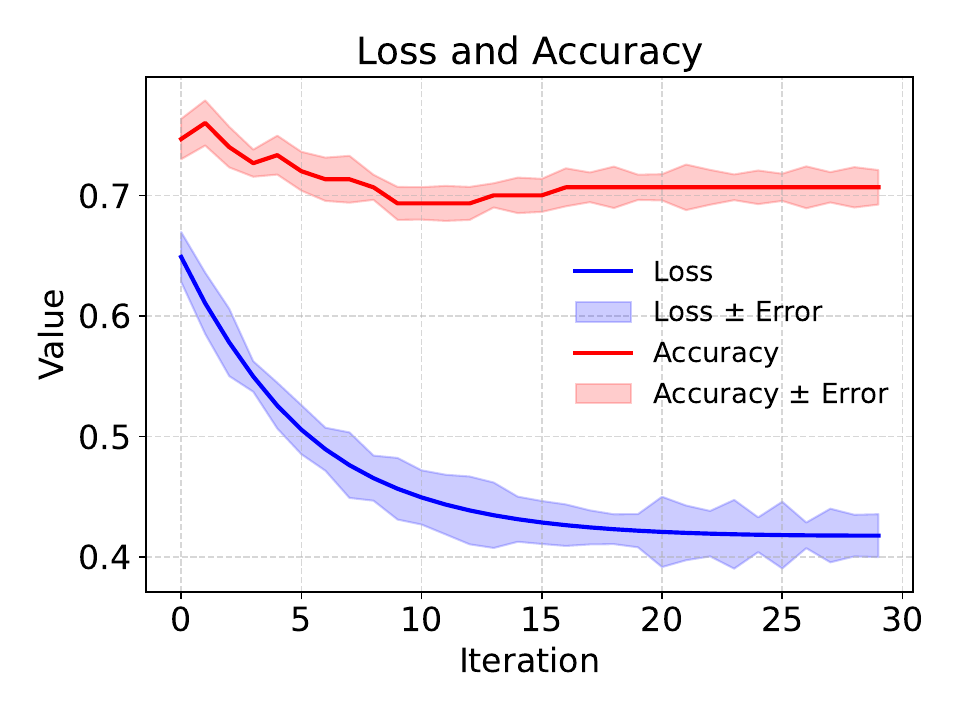}
        \caption{}
        \label{fig:gpsqnnqiskit}
    \end{subfigure}\hfill
    \begin{subfigure}{.33\textwidth}
        \includegraphics[width=\linewidth]{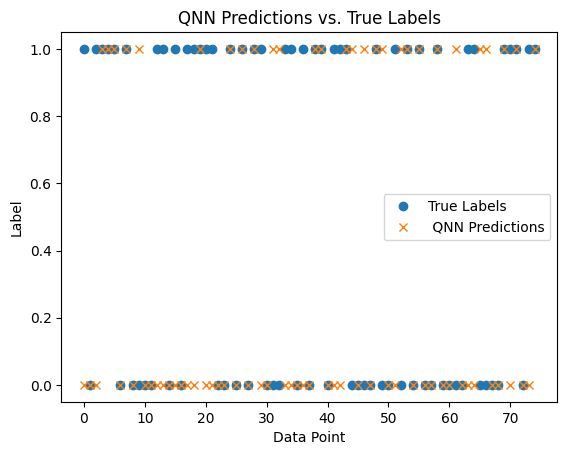}
        \caption{}
        \label{fig:gpsqnnqiskitdatapoints}
    \end{subfigure}
    \begin{subfigure}{.33\textwidth}
        \includegraphics[width=1\linewidth]{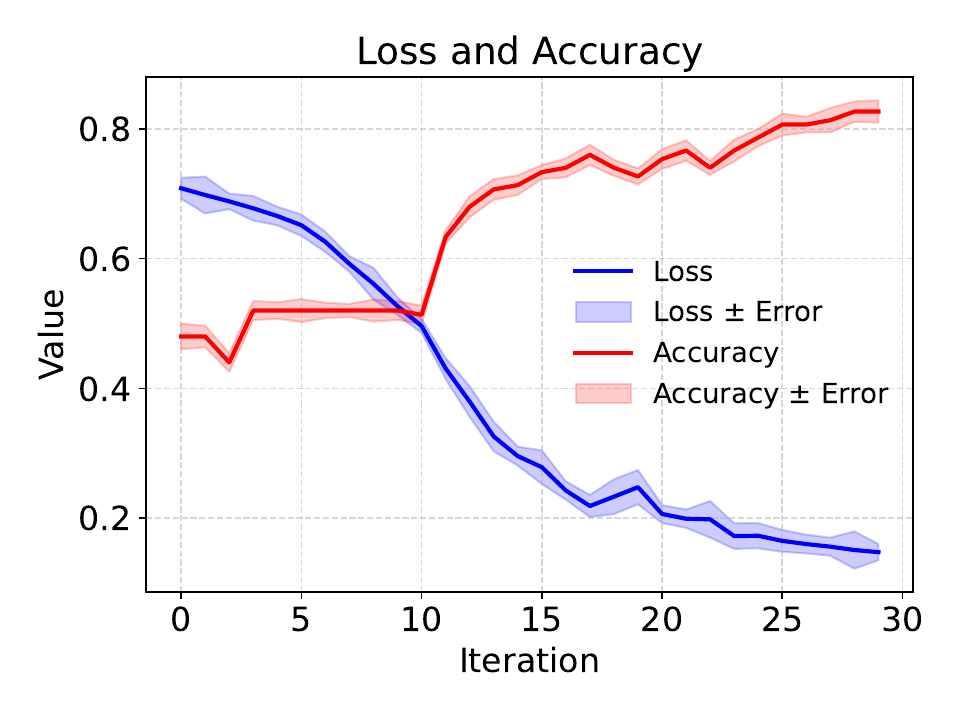}
        \caption{}
        \label{fig:gpshqnnqiskit}
    \end{subfigure}\hfill
    \begin{subfigure}{.33\textwidth}
        \includegraphics[width=\linewidth]{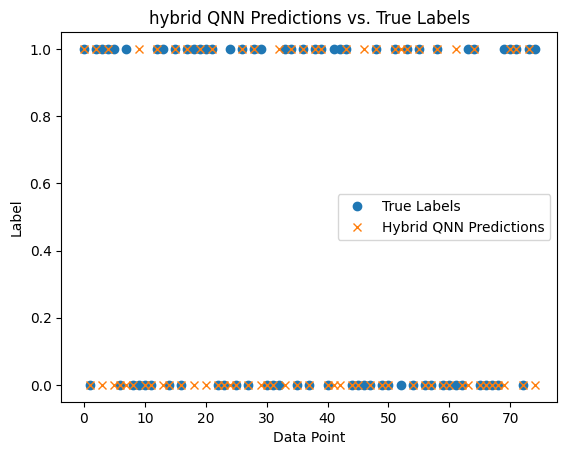}
        \caption{}
        \label{fig:gpshqnnqiskitdatapoints}
    \end{subfigure}
    \caption{Comparative results of different QNN approaches for GPSD. Loss and accuracy curves using (a) QNN-P, (c) QNN-Q, and (e) QNN-H. Predictions using (b) QNN-P, (d) QNN-Q, and (f) QNN-H.}
    \label{Fig5-}
\end{figure*}

\begin{figure}[htpb]
\centering
\begin{subfigure}{\linewidth}
\centering
\includegraphics[width=0.8\linewidth]{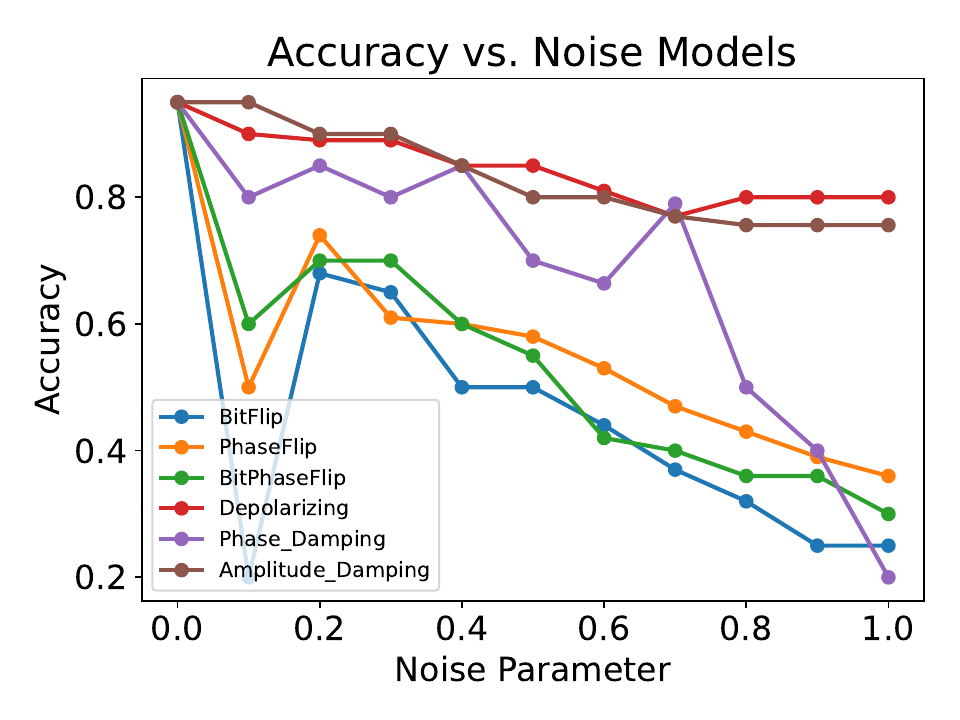}
\caption{}
\label{noise1}
\end{subfigure}\hfill
\begin{subfigure}{\linewidth}
\centering
\includegraphics[width=0.8\linewidth]{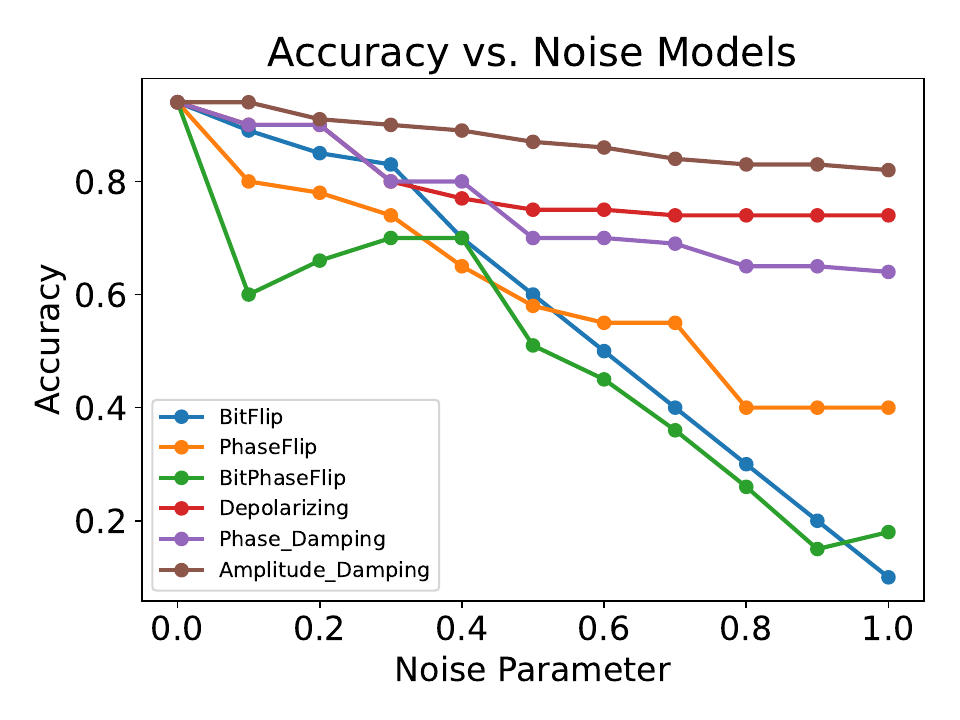}
\caption{}
\label{noise2}
\end{subfigure}\hfill
\caption{QNN-P Accuracy in the Presence of Six Noise Models for (a) RODD and (b) GPSD.}
\label{noise}
\end{figure}
Our research focuses on designing quantum and hybrid parameterized models using Qiskit and Pennylane and applying these models to multiple low-energy carbon IoT datasets to assess robustness and real-world applicability. Furthermore, it investigates how entanglement connections and circuit depth affect model performance. 

\subsection{Datasets}
The proposed quantum and hybrid models' performance is evaluated using two IoT datasets. The first, RODD \cite{D1}, includes features such as temperature, humidity, light, CO2, and humidity ratio, represented as decimal values. The second, GPSD {\cite{D2}, contains features like latitude and longitude and a ``type" attribute with descriptors such as ``normal," ``backdoor," ``DDoS," ``injection," and more. These datasets allow for testing model performance across different data types and scenarios.
\subsection{Preprocessing}
Preprocessing is a critical step in the modeling process. It begins with reducing the number of features using principal component analysis or feature selection, ensuring the focus is on the most relevant data. To address the class imbalance, data points are selected to represent labels 0 and 1 in the training set equally, which is essential for improving model performance on new data. Furthermore, the data is normalized with StandardScaler() to ensure consistent feature scaling followed by additional steps, including handling missing values, encoding categorical variables, managing outliers, and splitting the data into training and testing sets.
\subsection{Metrics and Hyperparameters}

The performance of the proposed algorithms is evaluated using the metrics of precision, recall, F1 score, and test accuracy. Precision quantifies the proportion of true positive predictions among all positive predictions, recall measures the ability to identify actual positives, and the F1 score balances precision and recall, while accuracy reflects the percentage of correctly predicted samples. The Binary Cross Entropy with Logits Loss (BCEWithLogitsLoss) function is used as the loss function. The training process employs the Adam optimizer, which iteratively updates model parameters across multiple epochs. Key hyperparameters for Adam include the learning rate and the exponential decay rates for moment estimates ($\beta_1 = 0.9$ {and} $\beta_2 = 0.999$), fine-tuned to ensure stable convergence. Fine-tuning the QNN-H architecture involves exploring various neural network configurations, such as the number of layers, neurons per layer, and activation functions (e.g., ReLU, Sigmoid, and Tanh). For quantum models, different embedding and architectures are explored and tested systematically until the final ones are selected, as described in the methodology.
\subsection{Results}
The evaluation of various QNN models for both the RODD and GPSD tasks demonstrates that the Pennylane implementation of QNN-P consistently achieves the highest performance as shown in Fig. \ref{Fig4-}. For the RODD task, QNN-P outperformed other models (Table \ref{Table2}), achieving a precision of 0.94, recall of 0.96, and F1-score of 0.95, reflecting its effectiveness in accurately identifying true positives while minimizing false positives. Additionally, its strong consistency between training accuracy (0.95) and test accuracy (0.90) suggests good generalization to unseen data. Compared to Qiskit-based QNN-Q and hybrid QNN-H models, QNN-P showed significant improvement, particularly in recall (0.96 vs 0.68) and F1-score (0.95 vs 0.73).

Similarly, for the GPSD, QNN-P demonstrated the best performance with a precision of 0.93, recall of 0.94, and F1-score of 0.95 (Table~\ref{dataset1}). The close alignment between training and test accuracy (both 0.94) underscores its robustness in handling new data. In contrast, QNN-Q displayed a significant gap between precision (0.68) and recall (0.816), leading to a lower F1-score of 0.741, while QNN-H achieved a more balanced but still lower performance, with a precision of 0.812, recall of 0.844, and F1-score of 0.827 (Fig. \ref{Fig5-}). These findings emphasize Pennylane's potential as a powerful framework for building high-performing QNNs, suggesting further exploration of its advantages for similar tasks.


\begin{table}[htbp]
\caption{RODD Performance Metrics.}
\label{Table2}
\centering
\footnotesize  
\setlength{\tabcolsep}{3pt} 
\renewcommand{\arraystretch}{1.1} 
\begin{tabularx}{\columnwidth}{@{}l *{6}{>{\centering\arraybackslash}X}@{}}  
\toprule
Model & Precision & Recall & F1 Score & Test Acc. & Runtime\\
\midrule
QNN-P & 0.94 & 0.96 & 0.95 & 0.95 & 1 min \\
QNN-Q & 0.81 & 0.68 & 0.73 & 0.80 & 7 min 40s\\
QNN-H & 0.93 & 0.88 & 0.90  & 0.91& 8 min 21s \\
SVM - Linear Kernel & 0.89 &  0.81 & 0.85 & 0.85 & 0.42s\\
Logistic Regression & 0.86 & 0.82 & 0.84 & 0.84 &  0.01s \\
Linear Discriminant Analysis & 0.87 & 0.83 & 0.85 & 0.85 &  0.0084s\\
Ridge Classifier & 0.87 & 0.83 & 0.85 & 0.85 & 0.0085s \\
Quadratic Discriminant Analysis & 0.84 & 0.83 & 0.84 & 0.84 & 0.05s \\
Navier Bayes & 0.86 & 0.83 & 0.84 & 0.84 & 0.006s \\
Gradient Boosting Classifier & 0.90 & 0.94 & 0.92 & 0.92 & 0.62s \\
Random Forest Classifier & 0.96 & 0.97 & 0.97 & 0.97 & 0.582s \\
Decision Tree Classifier & 0.95 & 0.96 & 0.96 & 0.95 & 0.03s \\
\bottomrule
\end{tabularx}
\end{table}
\begin{table}[htbp]
\caption{GPSD Performance Metrics.}
\label{dataset1}
\centering
\footnotesize  
\setlength{\tabcolsep}{3pt} 
\renewcommand{\arraystretch}{1.1} 
\begin{tabularx}{\columnwidth}{@{}l *{6}{>{\centering\arraybackslash}X}@{}}  
\toprule
Model & Precision & Recall & F1 Score & Test Acc. & Runtime \\
\midrule
QNN-P & 0.93 & 0.94 & 0.95  & 0.94 & 1 min 15s  \\
QNN-Q & 0.68 & 0.82 & 0.74 & 0.74 &  4 min 52s\\
QNN-H & 0.81 & 0.84 & 0.83 & 0.87 & 5 min 34s\\
SVM - Linear Kernel & 0.7646 &  0.6303 & 0.6910 & 0.71 & 3 min 22s\\
Logistic Regression & 0.76 & 0.62 & 0.68 & 0.71 &  0.07s \\
Linear Discriminant Analysis & 0.70 & 0.76 & 0.58 & 0.66 &  0.06s \\
Ridge Classifier & 0.76 & 0.58 & 0.66 & 0.70 & 0.05s \\
Quadratic Discriminant Analysis & 0.75 & 0.53 & 0.62 & 0.67 & 0.04s \\
Navier Bayes & 0.74 & 0.53 & 0.62 & 0.68 & 0.06s \\
Gradient Boosting Classifier & 0.83 & 0.94 & 0.88 & 0.88 & 11.53s \\
Random Forest Classifier & 0.96 & 0.95 & 0.96 & 0.96 & 18.36s \\
Decision Tree Classifier & 0.94 & 0.95 & 0.94 & 0.94 & 0.71s \\

\bottomrule
\end{tabularx}
\end{table}

\subsection{Noise Effect}

The comprehensive analysis of quantum noise models on the RODD and GPSD datasets using QNNs reveals critical insights into the resilience and vulnerabilities of QC systems under noisy conditions. The noise models were evaluated and tested across a noise parameter spectrum from 0 to 1. The impact of various quantum noise models on the performance of QNN-P for RODD and GPSD datasets is shown in Fig.~\ref{noise}. QNN-P consistently outperforms QNN-Q and QNN-H across multiple metrics for both datasets. This made QNN-P the most promising candidate for a detailed noise analysis. While QNN-Q and QNN-H were initially considered for analysis, their comparatively lower performance indicated that they would contribute less significant insights to this study. By focusing on QNN-P, we optimized resource allocation and gained a deeper understanding of noise behaviors.

For both datasets, the depolarizing and amplitude damping models exhibited superior resilience. The depolarizing model maintained consistent performance with only a gradual decrease in accuracy as the noise parameter increased, demonstrating its ability to preserve quantum state integrity despite environmental noise. The amplitude damping model, which reflects energy dissipation, also maintained commendable accuracy, suggesting its relevance in environments where energy loss occurs. In contrast, the bit flip, phase flip, and bit phase flip models were highly sensitive to noise, with significant drops in accuracy even at low noise levels. The bit flip model, in particular, showed a sharp decline in performance, underscoring the challenge of bit errors in QC. The phase damping model showed moderate resistance, simulating the loss of quantum information without energy loss, but still faced challenges in maintaining quantum coherence over time.
This analysis highlights the complex impact of different quantum noise types on QNN-P performance. The findings suggest that specialized quantum error correction strategies are necessary to improve reliability. Depolarizing and amplitude damping models are more suited for noisy environments, while bit flip, phase flip, and bit phase flip models require careful handling and robust error mitigation techniques.
\section{Discussion and Conclusion \label{SecV}}
This study explores the application of QNN models within the Pennylane and Qiskit frameworks, alongside the hybrid QNN-H, on two IoT datasets. It investigates how quantum algorithm settings, such as repetitions and entanglement connections, interact with the datasets, highlighting the significant influence of feature map and ansatz configurations on model performance. The QNN-P model demonstrates strong generalization, achieving 94\% and 95\% accuracy on the GPSD and RODD datasets, outperforming several classical models.
On the aspect of computational complexity and runtime, we tested various qubit numbers and circuit layers to achieve an optimal architecture that balances accuracy and computational constraints. Specifically, reducing the depth of quantum circuits without compromising model performance and minimizing gate operations and computational overhead to make the approach more suitable for the NISQ era (Tables \ref{Table2} and\ref{dataset1}). These metrics are calculated for 300 data points. Our sustainability approach focuses on employing minimal quantum circuit architectures to reduce computational complexity compared to larger circuits or classical methods. These optimized architectures require fewer qubits and shallower circuits, leading to fewer gate operations and shorter execution times, thereby decreasing energy consumption on quantum hardware. Additionally, classical simulations of these optimized architectures experience reduced computational overhead, resulting to lower overall energy usage. By designing quantum and hybrid architectures with minimal resource requirements, we align with the principles of sustainable computing and enhance energy efficiency.

To evaluate scalability, preliminary tests with reduced subsets of the datasets demonstrated consistent or slightly improved performance, underscoring the robustness of our models even when fewer data points were used. This indicates the potential of QNNs to scale effectively within the limitations of current datasets.

Deploying quantum solutions in real-world IoT systems faces challenges like hardware limitations, high costs, and infrastructure needs, including stable environments and access to QPUs. Integration with IoT devices also requires addressing compatibility issues, low-latency real-time processing, and energy efficiency concerns for resource-constrained devices.
In IoT-based smart cities and low-carbon environments, QNN-P’s resilience in noisy environments and its ability to generalize across datasets makes it suitable for tasks like energy optimization, traffic management, and resource distribution. These efficiencies can reduce the computational load on traditional systems, thus minimizing the carbon footprint of IoT applications. Moreover, QNNs could improve real-time decision-making in green technologies, such as renewable energy systems, by processing large datasets more effectively, leading to smarter energy usage and less waste. Future research aims to integrate tensor networks and implement quantum error mitigation techniques to improve quantum model effectiveness against noise further. Furthermore, incorporating larger datasets to explore scalability further as advancements in quantum hardware enable more complex computations.

\bibliographystyle{IEEEtran}
\bibliography{IEEEabrv,Bibliography}

\begin{thebibliography}{10}
\providecommand{\url}[1]{#1}
\csname url@rmstyle\endcsname
\providecommand{\newblock}{\relax}
\providecommand{\bibinfo}[2]{#2}
\providecommand\BIBentrySTDinterwordspacing{\spaceskip=0pt\relax}
\providecommand\BIBentryALTinterwordstretchfactor{4}
\providecommand\BIBentryALTinterwordspacing{\spaceskip=\fontdimen2\font plus
\BIBentryALTinterwordstretchfactor\fontdimen3\font minus \fontdimen4\font\relax}
\providecommand\BIBforeignlanguage[2]{{%
\expandafter\ifx\csname l@#1\endcsname\relax
\typeout{** WARNING: IEEEtran.bst: No hyphenation pattern has been}%
\typeout{** loaded for the language `#1'. Using the pattern for}%
\typeout{** the default language instead.}%
\else
\language=\csname l@#1\endcsname
\fi
#2}}

\bibitem{Dudhe2021IoT}
P.~Dudhe, N.~Kadam, R.~M. Hushangabade, and M.~S. Deshmukh, ``Internet of things (iot): An overview and its applications,'' \emph{2017 International Conference on Energy, Communication, Data Analytics and Soft Computing (ICECDS)}, 2018.

\bibitem{farooq2020role}
M.~S. Farooq, S.~Riaz, A.~Abid, T.~Umer, and Y.~B. Zikria, ``Role of iot technology in agriculture: A systematic literature review,'' \emph{Electronics}, vol.~9, no.~2, p. 319, 2020.

\bibitem{ihirwe2021cloud}
F.~Ihirwe, A.~Indamutsa, D.~Di~Ruscio, S.~Mazzini, and A.~Pierantonio, ``Cloud-based modeling in iot domain: a survey, open challenges and opportunities,'' in \emph{2021 ACM/IEEE International Conference on Model Driven Engineering Languages and Systems Companion (MODELS-C)}.\hskip 1em plus 0.5em minus 0.4em\relax IEEE, 2021, pp. 73--82.

\bibitem{zantalis2019review}
F.~Zantalis, G.~Koulouras, S.~Karabetsos, and D.~Kandris, ``A review of machine learning and iot in smart transportation,'' \emph{Future Internet}, vol.~11, no.~4, p.~94, 2019.

\bibitem{islam2020development}
M.~M. Islam, A.~Rahaman, and M.~R. Islam, ``Development of smart healthcare monitoring system in iot environment,'' \emph{SN computer science}, vol.~1, pp. 1--11, 2020.

\bibitem{boyes2018industrial}
H.~Boyes, B.~Hallaq, J.~Cunningham, and T.~Watson, ``The industrial internet of things (iiot): An analysis framework,'' \emph{Computers in industry}, vol. 101, pp. 1--12, 2018.

\bibitem{Karmakar2019IoT-IIoT}
A.~Karmakar, N.~Dey, T.~Baral, M.~Chowdhury, and M.~Rehan, ``Industrial internet of things: A review,'' \emph{2019 International Conference on Opto-Electronics and Applied Optics (Optronix)}, 2019.

\bibitem{raj2021predictive}
A.~Raj, S.~Kumar, and S.~Krishnan, ``Predictive maintenance in smart manufacturing using machine learning-based sensor data analysis,'' \emph{IEEE Internet of Things Journal}, vol.~8, pp. 2893--2902, 2021.

\bibitem{network1030017}
\BIBentryALTinterwordspacing
L.~Farhan, R.~S. Hameed, A.~S. Ahmed, A.~H. Fadel, W.~Gheth, L.~Alzubaidi, M.~A. Fadhel, and M.~Al-Amidie, ``Energy efficiency for green internet of things (iot) networks: A survey,'' \emph{Network}, vol.~1, no.~3, pp. 279--314, 2021. [Online]. Available: \url{https://www.mdpi.com/2673-8732/1/3/17}
\BIBentrySTDinterwordspacing

\bibitem{liao2018industrial}
Y.~Liao, E.~d. F.~R. Loures, and F.~Deschamps, ``Industrial internet of things: A systematic literature review and insights,'' \emph{IEEE Internet of Things Journal}, vol.~5, no.~6, pp. 4515--4525, 2018.

\bibitem{Ahmed2023IIoT}
S.~F. Ahmed, M.~S.~B. Alam, M.~Hoque, A.~Lameesa, S.~Afrin, T.~Farah, M.~Kabir, G.~Shafiullah, and S.~Muyeen, ``Industrial internet of things enabled technologies, challenges, and future directions,'' \emph{Computers and Electrical Engineering}, vol. 110, p. 108847, 2023.

\bibitem{sisinni2018industrial}
E.~Sisinni, A.~Saifullah, S.~Han, U.~Jennehag, and M.~Gidlund, ``Industrial internet of things: Challenges, opportunities, and directions,'' \emph{IEEE transactions on industrial informatics}, vol.~14, no.~11, pp. 4724--4734, 2018.

\bibitem{khan2020industrial}
W.~Z. Khan, M.~Rehman, H.~M. Zangoti, M.~K. Afzal, N.~Armi, and K.~Salah, ``Industrial internet of things: Recent advances, enabling technologies and open challenges,'' \emph{Computers \& electrical engineering}, vol.~81, p. 106522, 2020.

\bibitem{rahman2021quantum}
A.~Rahman, X.~Li, and Y.~Chen, ``Quantum computing for optimizing energy efficiency in iiot: A comprehensive review,'' \emph{IEEE Internet of Things Journal}, vol.~8, pp. 5605--5617, 2021.

\bibitem{smith2020energy}
J.~Smith, A.~Patel, and W.~Chang, ``Energy-efficient solutions for iiot using quantum-inspired algorithms,'' \emph{IEEE Internet of Things Journal}, vol.~7, pp. 4751--4760, 2020.

\bibitem{bhatia2023novel}
M.~Bhatia, S.~Sood, and V.~Sood, ``A novel quantum-inspired solution for high-performance energy-efficient data acquisition from iot networks,'' \emph{Journal of Ambient Intelligence and Humanized Computing}, vol.~14, no.~5, pp. 5001--5020, 2023.

\bibitem{kaur2022generative}
I.~Kaur, E.~L. Lydia, V.~K. Nassa, B.~Shrestha, J.~Nebhen, S.~Malebary, and G.~P. Joshi, ``Generative adversarial networks with quantum optimization model for mobile edge computing in iot big data,'' \emph{Wireless Personal Communications}, pp. 1--21, 2022.

\bibitem{raparthi2022quantum}
M.~Raparthi, ``Quantum-inspired optimization techniques for iot networks: Focusing on resource allocation and network efficiency enhancement for improved iot functionality,'' \emph{Advances in Deep Learning Techniques}, vol.~2, no.~2, pp. 1--9, 2022.

\bibitem{emu2022resource}
M.~Emu, S.~Choudhury, and K.~Salomaa, ``Resource optimization of sfc embedding for iot networks using quantum computing,'' in \emph{2022 IEEE 27th International Workshop on Computer Aided Modeling and Design of Communication Links and Networks (CAMAD)}.\hskip 1em plus 0.5em minus 0.4em\relax IEEE, 2022, pp. 83--88.

\bibitem{satpathy2023analysis}
S.~K. Satpathy, V.~Vibhu, B.~K. Behera, S.~Al-Kuwari, S.~Mumtaz, and A.~Farouk, ``Analysis of quantum machine learning algorithms in noisy channels for classification tasks in the iot extreme environment,'' \emph{IEEE Internet of Things Journal}, 2023.

\bibitem{D1}
\BIBentryALTinterwordspacing
V.~F. Luis M.~Candanedo, ``Room occupancy detection data (iot sensor).'' [Online]. Available: \url{https://www.kaggle.com/datasets/kukuroo3/room-occupancy-detection-data-iot-sensor}
\BIBentrySTDinterwordspacing

\bibitem{D2}
\BIBentryALTinterwordspacing
N.~Moustafa, ``Ton\_iot datasets.'' [Online]. Available: \url{https://research.unsw.edu.au/projects/toniot-datasets}
\BIBentrySTDinterwordspacing

\end{thebibliography}
\end{document}